\documentclass[pra,showpacs,twocolumn,preprintnumbers,amsmath,amssymb,superscriptaddress]{revtex4}

\usepackage{graphicx}
\usepackage{dcolumn}
\usepackage{bm}

\begin{document}


\title{Steady-state generation of negative-Wigner-function light using feedback}

\author{Cristian Joana}
\affiliation{National Institute of Informatics, 2-1-2 Hitotsubashi, Chiyoda-ku, Tokyo 101-8430, Japan}

\author{Peter van Loock}
\affiliation{Institute of Physics, Johannes Gutenberg-Universit\"{a}t Mainz, 55099 Mainz, Germany}

\author{Hui Deng}
\affiliation{Department of Physics, University of Michigan, Ann Arbor, MI 48109, USA} 

\author{Tim Byrnes}
\affiliation{New York University Shanghai, 1555 Century Ave, Pudong, Shanghai 200122, China}
\affiliation{NYU-ECNU Institute of Physics at NYU Shanghai, 3663 Zhongshan Road North, Shanghai 200062, China}
\affiliation{National Institute of Informatics, 2-1-2 Hitotsubashi, Chiyoda-ku, Tokyo 101-8430, Japan}
\affiliation{Department of Physics, New York University, New York, NY 10003, USA}

\date{\today}

\begin{abstract}
We propose a method of producing steady-state coherent light with negative Wigner functions in nonlinear media combined with feedback control. While the nonlinearities are essential to produce the Wigner negativities, this alone is insufficient to stabilize steady-state light with negativities. Using feedback control to control the phase in the cavity we find that this produces significant total negativities for reasonable experimental parameters. The negative Wigner function is produced continuously and does not appear to be restricted to low amplitude light.   The technique is applicable to systems such as exciton-polaritons, where strong natural nonlinearities are present. 
\end{abstract}


\maketitle

\section{Introduction}
Producing and analyzing nonclassical states of light has been one of the long-standing goals in the field of quantum optics 
\cite{IOPNCSIQP,ToNCSoL,WallsMilburn,ScienceQTl}. In the context of quantum metrology  and information there has been a large amount of activity relating to the generation of nonclassical states of light such as squeezed states, single photon states, NOON states and Schrodinger cat/kitten states \cite{PhysRevX3041028,PhysRevA81033838,Ast:12,NatureN00N,Sciencekitten}.
While squeezed states are routinely producible in the laboratory, other types of nonclassical light are not as easily produced.  Specifically, non-Gaussian light, for example those with a negative component to the Wigner distribution, are significantly more difficult to produce. While in principle sending coherent light through a nonlinear Kerr medium (such as an optical fiber) will generate such light, the strength of the nonlinearity compared to the loss rates make this approach difficult in practice 
 \cite{stobinska08}. Methods to enhance the nonlinearity using various means have been investigated \cite{2008NatPh4890M,stob18,PhysRevA82}, but still difficulties remain before such methods are adapted towards a practical light source.

Some other methods to produce light with negative Wigner distributions include photon subtraction \cite{PRL97,PhysRevLett98},  
single (and more generally Fock state) photon generation \cite{PhysRevLett96}.
For photon subtraction methods, the nonlinearity originates from the measurement process and can produce highly non-Gaussian light. The price to be paid for the nonlinearity is that the method is inherently probabilistic, thus as a source of non-Gaussian light it is limited in efficiency and cannot be used easily in a continuous wave setting. Single and Fock state photon generation also produce highly non-Gaussian light, but are generally at the few-photon level, and hence are typically much darker than photon subtraction methods.  The probabilistic aspect also enters in such schemes  even for deterministic sources, as efficiency issues are present which greatly reduce the generation probability from unity \cite{RevSciInstrum82}. Therefore, there is currently no method that is available to produce light with a negative Wigner component at steady-state, furthermore beyond the few-photon level.

A parallel development in the field of semiconductor optics has been the realization of exciton-polariton Bose-Einstein condensation (BEC) \cite{deng10,kavokin10,richard10,snoke10,keeling11,timofeev12,carusotto13,byrnes14}. 
Exciton-polaritons are quasiparticle excitations  formed in planar semiconductor microcavity structures in a superposition of an exciton (an electron-hole bound pair) and a cavity photon.  They have an effective mass that is extremely light -- inherited from the photon component -- and typically is $ 10^{-4} $ times the bare electron mass.  Their matter component produces a polariton-polariton interaction that originates from a Coulomb exchange \cite{PhysRevB81205312,PhysRevB587926}.
These properties of exciton-polaritons have allowed for the observation of Bose-Einstein condensation in such semiconductor microstructures \cite{kasprzak06,balili07,deng02}. Most exciton-polariton BEC experiments have been conducted at typically $ \sim 10$ K in CdSe and GaAs based samples.  However, room temperature condensation has also been observed experimentally using other materials such as GaN, ZnO, and organic semiconductors \cite{christopoulos07,baumberg08,kenacohen10,guillet11,plumhof14}. This makes them an attractive system for future technological applications.  

\begin{figure}[t]
\includegraphics[width=0.9\columnwidth]{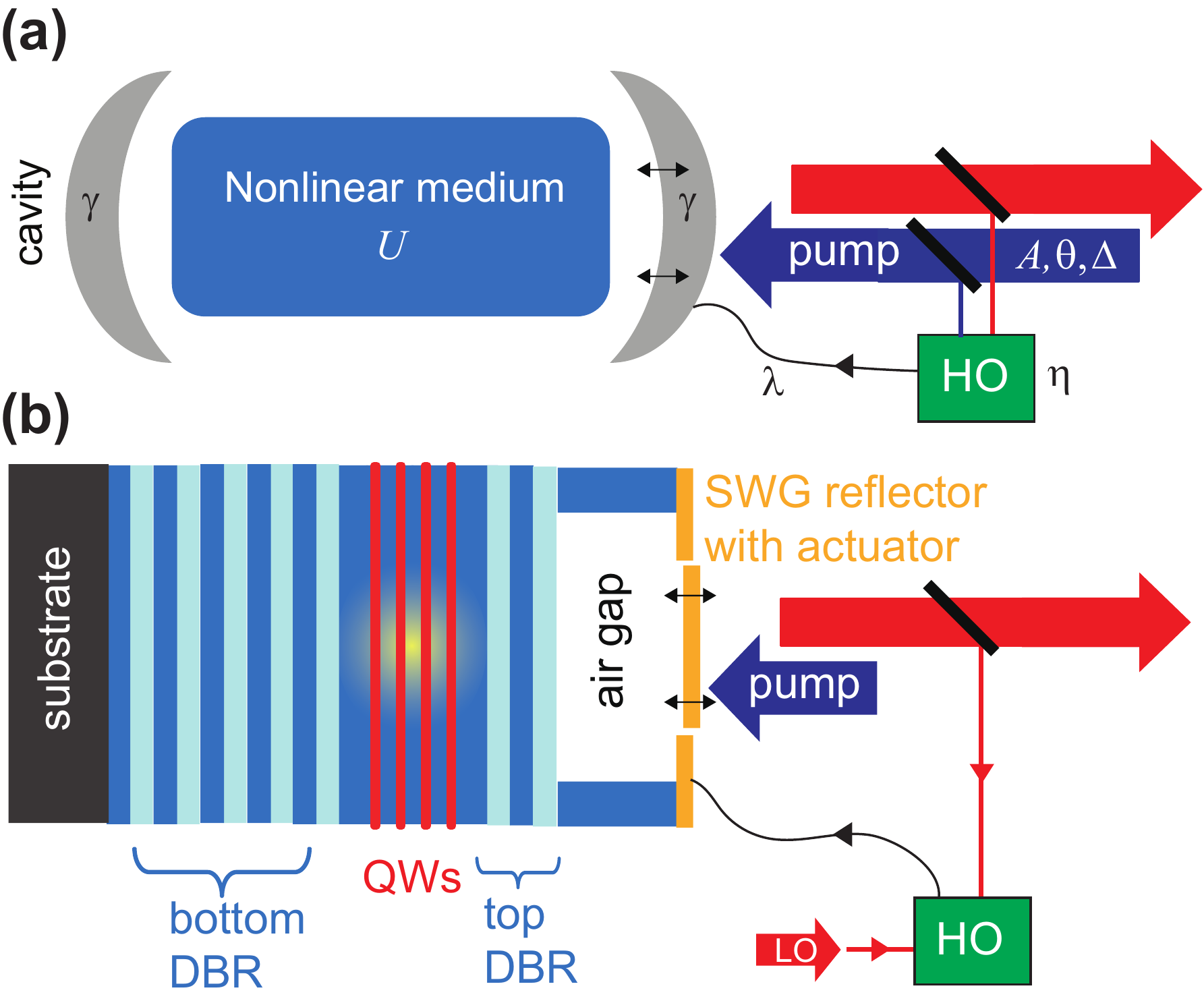}
\caption{\label{fig1}(Color online) Experimental scheme considered in this paper.  (a) The basic description of the ingredients necessary for steady-state negative Wigner function light generation. The parameters as defined in the master equation (\ref{master}) are labeled. (b) The specific implementation with exciton-polaritons.  A semiconductor microcavity structure with quantum wells (QWs) allows for the excitation of exciton-polaritons by the pump laser. The microcavity is formed by a combination of distributed Bragg reflectors (DBRs) on either side of the QWs, and a floating sub-wavelength grating (SWG) reflector. The resonant frequency of the cavity can be controlled by nanoelectromechanical actuators, in a way such as that realized in Ref. \cite{huang08}. The phase of the output light is fixed with reference to a phase-stable local oscillator (LO) and is determined by homodyne detection.}
\end{figure}

Such exciton-polaritons naturally emit coherent light, which is inherited from the coherence of the BEC. The light emerges vertically through the microcavity mirrors as a result of the finite lifetime of the polaritons.  The quality of the cavities has improved with fabrication techniques -- in the first experiments lifetimes of the polaritons were in the region of $ \sim 1 $ ps \cite{kasprzak06,deng02}, however lifetimes of $ \sim 10-100 $ ps are now realizable \cite{nelsen13,tanese13}. Comparing the energy scales of the polariton-polariton interaction with the cavity decay, these improvements put the two processes on a similar order. In laterally confined systems, the interaction energy of polaritons in the lasing mode has exceeded the cavity decay \cite{huiref}. This should be compared to typical nonlinear media where the decay rate is larger than the nonlinear coefficient by a factor of $ \sim 10^5 $ \cite{munro05}. This, and the potential room-temperature operation, makes the devices attractive as a potential way of generating coherent non-Gaussian light.

Previously we introduced a method based on exciton-polariton BECs to generate bright, steady-state non-Gaussian 
light \cite{byrnes13}. In the scheme, a pump laser produces a hot gas of exciton-polaritons, which condenses into the zero momentum state due to polariton-polariton scattering.  At sufficiently high densities, a macroscopic population of polaritons develops in the condensate, 
and with suitable phase fixing methods, we showed that non-Gaussian light could be generated at steady-state. 
 Our aim in this paper is to generalize the method to any system possessing a non-linearity in combination with feedback.  In Ref. \cite{byrnes13} we used a particular master equation that was specific to exciton-polaritons, but as we show in this paper, the stabilization of negative Wigner function light is possible with any non-linear medium in a cavity when combined with feedback. 

For the exciton-polariton implementation, we also provide several improvements upon the technique as described in Ref. \cite{byrnes13}.   
First, while in our previous work off-resonant pumping was used to pump the system, which provided the population of hot polaritons for the polariton condensate, we show that resonant pumping of the condensate directly is feasible. The indirect pumping method is typically used for studying polariton condensation; however, in our case we are interested in the coherent light that is emitted by the sample. Hence resonantly and directly exciting the zero momentum mode is adequate for our purposes.  The possibility of resonant pumping greatly simplifies the procedure from an experimental point of view as it is not necessary to generate conditions such that a BEC can form.  Second, we show that by using a feedback based phase locking for the condensate, it is possible to create highly non-Gaussian states with greatly improved negativities in the Wigner function. Our current technique improves the total negativity achievable by typically an order of magnitude for equivalent experimental parameters. 

This paper is organized as follows.  In Sec. \ref{sec:expmodel} we discuss the experimental setup required for realizing steady-state non-Gaussian light.  The model for the nonlinear dissipative system with feedback is described, with the numerical methods to solve the system.  In Sec. \ref{sec:wigner} we show the time evolutions of the Wigner functions towards steady-state.  In Sec. \ref{sec:negativity} we explore the available parameter space and discuss under what conditions the most negative Wigner functions are obtained.  In Sec. \ref{sec:conc} we summarize and discuss our conclusions.

\section{Proposed experimental system}
\label{sec:expmodel}

\subsection{Nonlinear cavity with feedback}

The experimental system that we consider in this paper is shown in Fig. \ref{fig1}(a). A $ \chi^{(3)} $ Kerr nonlinear medium of strength $ U $ is placed in a cavity with decay rate $ \gamma $, and pumped with coherent light with a displacement of $ A e^{i \theta} $.   The pump may have a detuning with respect to the cavity mode resonance by a frequency $ \Delta $.  The light emerging from the cavity is passed through a beamsplitter of reflectance $ r $, of which the reflected part is measured using a homodyne measurement. The result of the homodyne measurement is passed back to the cavity, where the cavity dynamics is modified in real time.  The type of feedback assumed here is the same as that modeled by Wiseman and Milburn in Ref. \cite{wiseman93}. 
The aim is to produce light which has Wigner negativities that are stabilized at steady-state.  As discussed in Ref. \cite{byrnes13}, a Kerr nonlinear medium will produce Wigner negativities transiently, but it is much more difficult to achieve this at steady-state due to phase diffusion effects from the cavity.  As we show in this paper, the feedback acts to stabilize this, and gives the capability to produce light with negative Wigner function characteristics at steady-state. 

A more specific experimental configuration using exciton-polaritons is shown in Fig.  \ref{fig1}(b). The basic configuration is that of a semiconductor quantum well (QW) placed within a microcavity in a strong coupling regime, forming exciton-polaritons.  In contrast to a standard exciton-polariton microcavity where both the mirrors are formed by distributed Bragg reflectors (DBRs) on either side of the QWs, we consider that on one side the reflector is a sub-wavelength grating (SWG) \cite{huang08}.  Strong coupling of exciton-polaritons in a SWG cavity has been realized; furthermore a polariton laser has also been realized in such cavities \cite{zhang14}. A polariton laser differs from a polariton BEC only in the sense of what level of thermalization is achieved \cite{byrnes14}.  Since we consider resonant excitation of the polariton condensate, the thermalization aspect is irrelevant for our purposes.  The advantage of the SWG is that the cavity can be modulated at high frequencies thanks to the light mass of the mirrors on the order of nanoseconds \cite{huang08}. In order to be effective for the feedback, this should be of a similar order or faster than the coherence times for polaritons.  This can be considerably longer than the lifetime of the polaritons themselves,  being also in nanosecond scale \cite{love08,Nat4648,Steger:15}.  The exciton-polaritons are excited coherently and resonantly with a pump laser in the zero in-plane momentum direction (perpendicular to the QWs).  As discussed above, it is essential to control the phase of the polariton condensate in order to take advantage of the intrinsic nonlinearity that the polaritons experience when in the QWs.  To achieve this, a homodyne measurement of the output light of the polaritons is performed against a local oscillator with a relatively stable phase \cite{199292}.  The positive $ x $ homodyne current is fed back to the SWG actuator.  This stabilizes the phase of the output light, which as we show below, results in steady-state nonclassical light with a negative Wigner function component.

\subsection{Master equation}

The system described above can be described by a master equation
\begin{align}
\begin{split}\label{master} 
\frac{d \rho}{dt} =  & \frac {i}{ \hbar} [H_0 + H_{\mbox{\tiny int}}+ H_{\mbox{\tiny pump}}, \rho ] - \frac{\gamma}{2}  \text{$\cal{L}$} [a, \rho\,]  \\
& - \frac{\lambda^2}{2 \eta} \big[a^\dagger a, [a^\dagger a, \rho]\big] + i \lambda [a^\dagger a,\rho a^\dagger + a \rho] ,
\end{split}
\end{align}
where the Hamiltonians are
\begin{align}
H_0 &  = \hbar \Delta a^\dagger a ,  \\
H_{\mbox{\tiny pump}} &  = \hbar A (e^{i\theta} a^\dagger   - e^{-i\theta} a ) ,  \\
H_{\mbox{\tiny int}} & = \frac{U}{2} a^\dagger a^\dagger a a    . \label{intham} 
\end{align}
Here, $ a, a^\dagger $ are the annihilation and creation operators for the photons inside the cavity; $H_0 $ is the detuning between the pump laser and cavity resonance; $H_{\mbox{\tiny int}}$ is the nonlinear photon interaction; $ H_{\mbox{\tiny pump}} $
is the coherent pump laser excitation with amplitude $ A $, and $ \theta $ is the phase difference between the pump laser and 
the feedback phase locking.  The superoperator 
\begin{equation}
 {\cal L}[a,\rho] \equiv a^\dagger a\rho + \rho a^\dagger a - 2 a\rho a^\dagger 
\end{equation}
is the Lindblad loss term for photons leaking through the cavity with rate $ \gamma $. The last two terms in Eq. (\ref{master}) describe the feedback phase stabilization via homodyne detection \cite{wiseman93}: $ \eta $ is the effective detection efficiency (see the Appendix for the definition), $ \lambda $ is the coefficient of the feedback process. The mode structure inside and outside the cavity is given in the Appendix.   

For the more specific case of exciton-polaritons, we simply re-interpret the above parameters for the microcavity structure.  Here, $ a, a^\dagger $ are the annihilation and creation operator for the zero momentum polaritons inside the QWs; $H_0 $ is the detuning between the pump laser and the zero momentum polariton energy; $H_{\mbox{\tiny int}}$ is the polariton-polariton interaction.  The pump laser directly pumps the polaritons at zero transverse momentum with amplitude $ A $, and phase $ \theta $.  We note that we make the standard assumption that the quantum statistics of the polaritons is the same as the light emerging from the microcavity, as this is a coherent process conserving momentum and energy \cite{byrnes14}.  This has been used to measure polariton correlation functions successfully \cite{horikiri10}. 

After evolving the density matrix decomposed in Fock states given by
\begin{equation}
\rho = \sum_{n,m} \rho_{nm} | n  \rangle \langle m | 
\end{equation}
we find the steady-state solutions which are  converted into a Wigner distribution function \cite{WallsMilburn,stobinska08} 
according to 
\begin{align}
W(\alpha,\alpha^*)= \frac{2 e^{2|\alpha|^2}}{\pi} \sum_{n,m} \frac{\rho_{nm}}{(-2)^{n+m}} 
\frac{ \partial^{n+m}}{\partial \alpha^{*n} \partial \alpha^{m}} e^{-4|\alpha|^2}  
\end{align}
A maximum photon number cutoff $ n_{\mbox{\tiny cut}} =100$ is used for all the simulations which is sufficient for convergence. 

Let us now estimate the parameters based on reasonable experimentally achievable parameters.  We will base these on those
achieved with exciton-polaritons,  which we consider to be a strong candidate for realizing the present scheme.  We choose two parameters sets corresponding to (i) the currently available SWG cavities and (ii) the best available high quality cavities (conventional top and bottom DBR, not SWG).  Current SWG based cavities have a polariton lifetime in the region of $ 1/\gamma \sim 6 $ ps, while the best available conventional polariton cavities are in the region of $ 1/\gamma \sim 100 $ ps \cite{nelsen13,Dreismann17062014}. However the lifetime of polaritons in SWG cavities can be improved if the grating design and fabrication are further refined, improving the lateral confinement. The interaction energy can be estimated from the formula
\begin{align}
U = \frac{30 e^2 a_B |X|^4}{\pi^3 \epsilon A}
\end{align}
where $ e $ is the electronic charge, $ X $ is the exciton Hopfield coefficient, $ \epsilon $ is the effective permittivity in the semiconductor, $ a_B $ is the Bohr radius, and $ A $ is the trapping area \cite{PhysRevB81205312,byrnes10}.  For GaAs based samples with $ a_B = 10 $ nm, zero detuning $ X = 1/\sqrt{2} $, a 1 $\mu$m diameter spot size, $ \epsilon = 13 \epsilon_0$, we obtain $ U \sim 4 \mu$eV.  This is in line with our experimental estimates of $ U $ in SWG based polariton cavities. We thus use $ U/\hbar \gamma \approx 0.05,0.5 $ for current SWG cavities and high-Q cavities respectively.

\begin{figure}[t]
\includegraphics[width=\columnwidth]{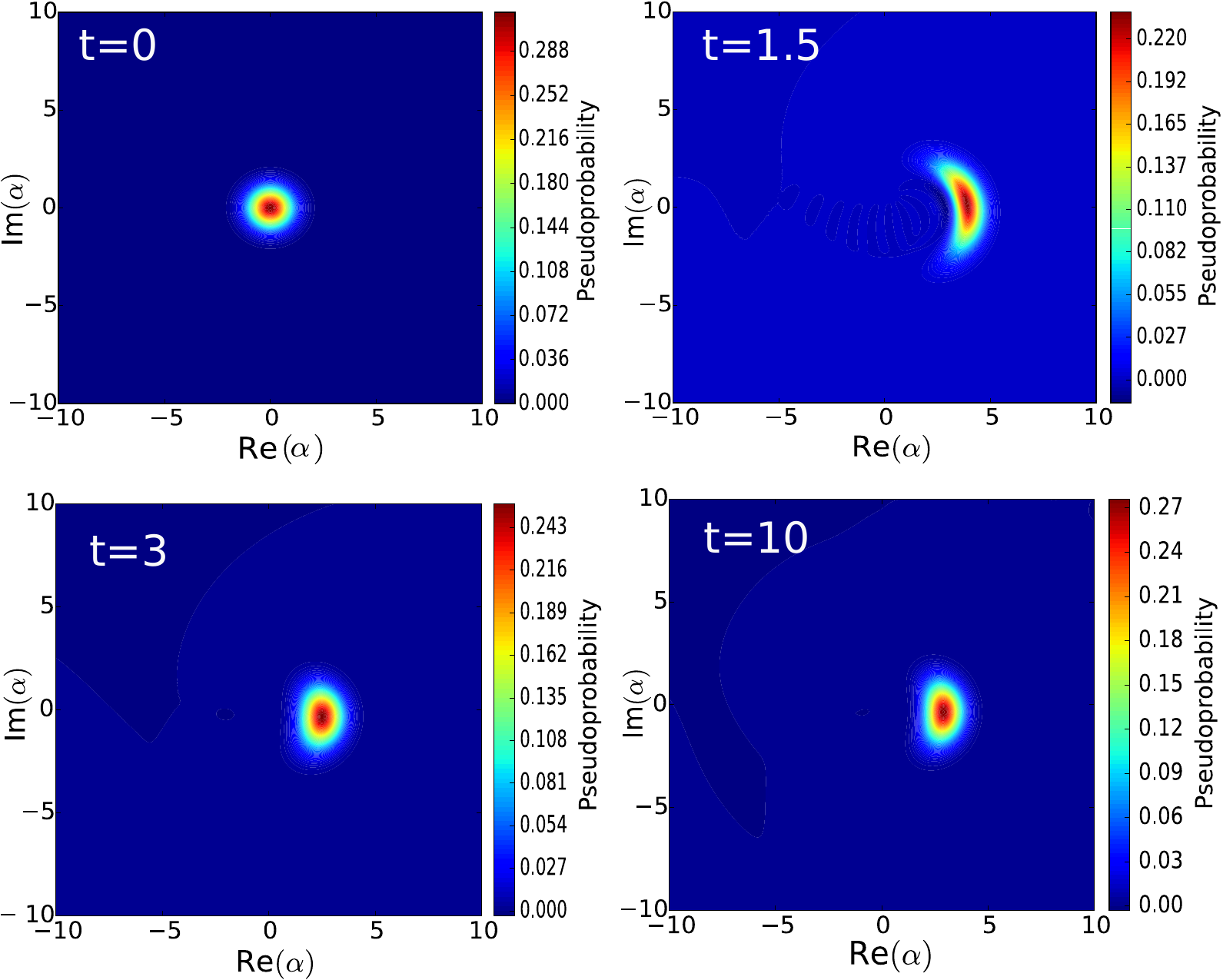}
\caption{\label{fig2}(Color online)  Time evolution of the  Wigner function distributions for coherently excited cavity photons (polaritons) with no feedback. The used parameters are $\lambda=0$, $U/\gamma\hbar=0.3$, $A=3$ and $\theta=60^\circ$.}
\end{figure}

\section{Time evolution of Wigner functions}
\label{sec:wigner}

We now show the results of the time evolution of the master equation (\ref{master}).  First we show our results without performing feedback on the system ($\lambda = 0 $).  Without interactions $ U = 0 $, we have a lossy system with a coherent pump, which results in a displaced coherent state.  The steady-state solution may be found by writing the evolution for the coherent state amplitude
\begin{align}
\frac{ d \langle a \rangle }{dt} = i \Delta \langle a \rangle  + i A e^{i\theta} - \frac\gamma2 \langle a \rangle. 
\end{align}
Writing the amplitude in the rotating frame,  with  $\alpha = \langle a e^{-i \Delta t}\rangle $, 
\begin{align}
\frac{ d \alpha }{dt}  =  i A e^{i(\theta-  \Delta t)} - \frac\gamma2 \alpha . 
\end{align}
Setting $ \frac{ d\langle a \rangle}{dt} = 0 $ gives a steady-state amplitude
\begin{align}
\alpha = \frac{i 2A e^{i(\theta-  \Delta t)} }{\gamma} .
\end{align}
The coherent state thus has an amplitude $ | \alpha | = 2A/ \gamma $ at a phase which depends on $ \theta $ and $ \Delta $.  

Introducing the nonlinearity $ U$, we have Wigner distributions which are shown in Fig. \ref{fig2}.  In the initial stages after the pump is turned on, the coherent evolves into a ``banana'' shape.  This is reminiscent of the time evolution of a coherent state when it is put in a Kerr nonlinearity \cite{stobinska08}.  In an ideal Kerr medium without losses, the banana shape then develops into ``ripples'' with negative regions in the Wigner function, followed by an evolution into a Schrodinger cat state.  In our case, the banana shape reverses and again becomes shorter, finally stabilizing into an approximately oval shape.  For some parameter ranges in the intermediate regime some negativity of the Wigner function is created.  However, for all parameter ranges no negativity is observed at steady-state.

\begin{figure}[t]
\includegraphics[width=\columnwidth]{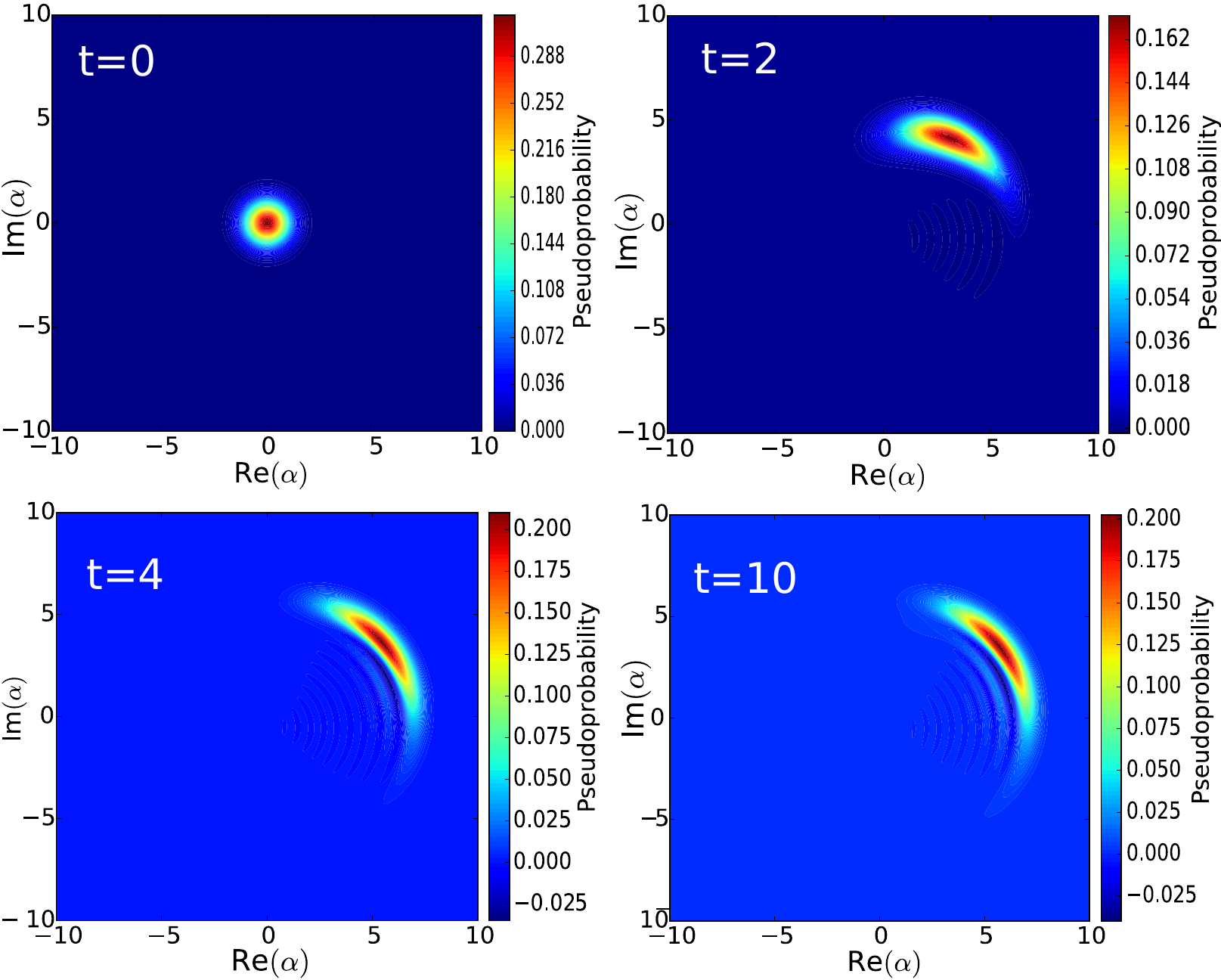}
\caption{\label{fig3}(Color online)   Time evolution of the  Wigner function distributions for coherently excited cavity photons (polaritons) with feedback.The used parameters are $\lambda=0.8$, $U/\gamma\hbar=0.3$, $A/\gamma=3$ and $\theta=60^\circ$.}
\end{figure}

When feedback is introduced, the situation changes dramatically (see Fig. \ref{fig3}).  Again, as time evolves, the Wigner function evolves into a banana shape due to the presence of the nonlinearity.  Unlike the feedback case where the Wigner function reverts to a oval shape, the feedback maintains the length of the banana, and steadily increases towards an equilibrium configuration.  Furthermore, the ripples in the vicinity of the banana which give negative regions to the Wigner function are preserved.  The preservation of the ripples allow for negativity to remain even at steady-state.

\begin{figure}[t]
\includegraphics[width=\columnwidth]{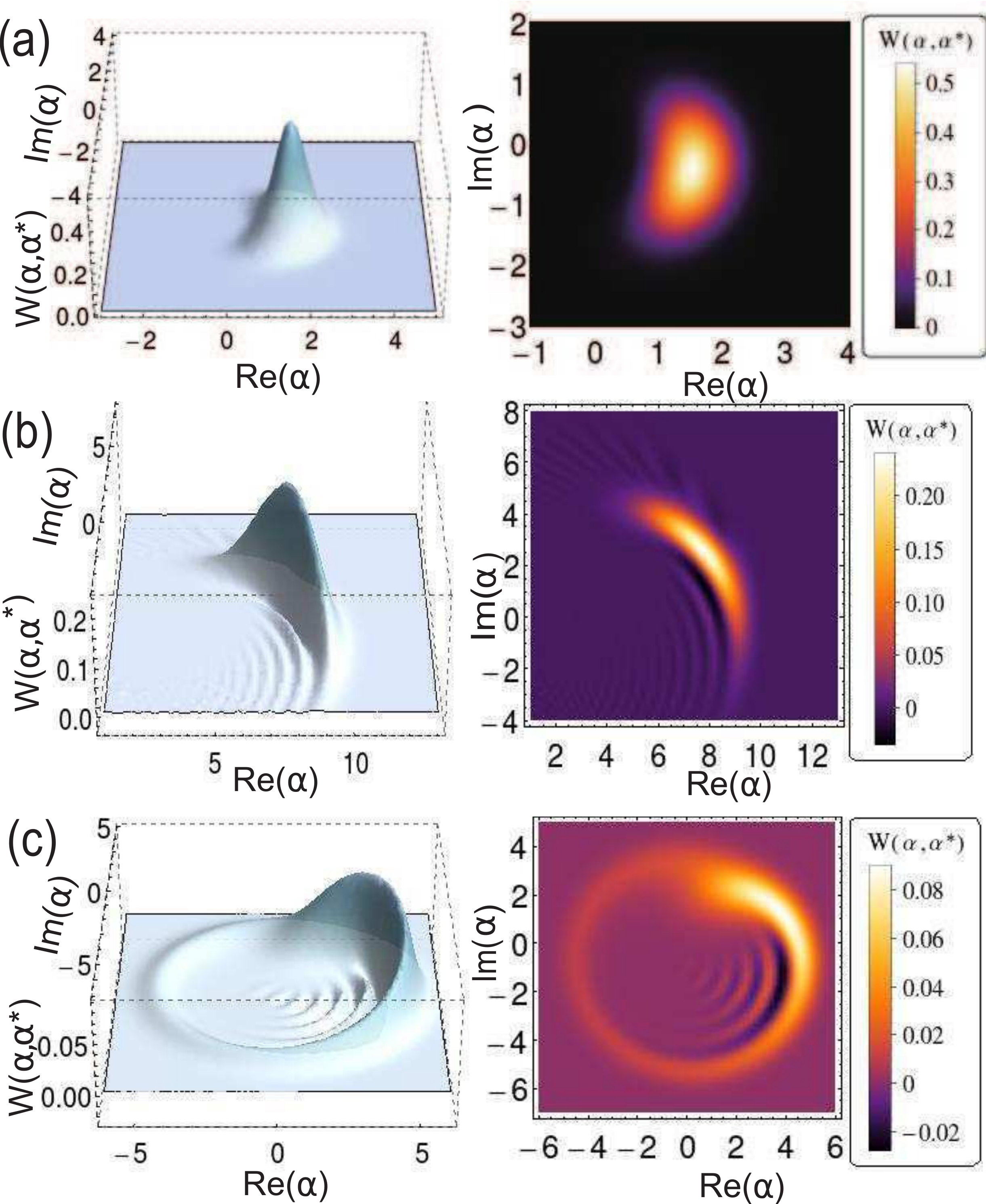}
\caption{\label{fig4}(Color online)  Steady-state Wigner function distributions for coherently excited cavity photons (polaritons).  (a) The case with no feedback: $ \lambda/\gamma =0 $, $ \Delta/\gamma=0 $, $ U/\hbar \gamma = 0.5 $, $ A/\gamma= 3 $, $\theta=0^\circ$, $\eta = 1 $; (b) typical distribution for a low-Q cavity: $ \lambda/\gamma = 0.16 $, $ \Delta/\gamma= 0 $, $ U/\hbar \gamma = 0.05 $, $ A/\gamma= 3 $, $\theta=10^\circ$, $\eta = 1 $; (c) high-Q cavity: $ \lambda/\gamma =0.65 $, $ \Delta/\gamma=0 $, $ U/\hbar \gamma = 0.5 $, $ A/\gamma= 3 $, $\theta=-5^\circ$, $\eta = 1.0$.}
\end{figure}

Figure \ref{fig4} shows steady-state results for a range of parameters. In Fig.  \ref{fig4}(a) we show results for no feedback, for the case where $ U \gg \hbar \gamma $.  Naively one might expect that for large enough $ U$ one would be able to produce Wigner functions with negative components. We see that even for a very large value of $ U $, this is not true and the Wigner function evolves to completely positive distributions. In such a situation it is possible to produce light with negative Wigner components transiently, but as $ t \rightarrow \infty $, the Wigner distributions become always positive definite.  

What is the reason for this lack of negativity in the steady-state? We attribute this to the fact that despite the presence of nonlinear interactions (\ref{intham}), in the Fock basis, this produces a relative phase between different Fock states.  
The cavity introduces an effective phase diffusion term where the phase of the condensate randomly drifts with time.  This acts to completely destroy any phase that is generated by (\ref{intham}) in the steady-state regime.  Another way to understand this is that at steady-state the effect of phase diffusion is to reduce a coherent state to a mixture of coherent states with all phases \cite{rudolph01}
\begin{align}
\rho & = \frac{1}{2\pi} \int_0^{2 \pi} d \phi | \alpha e^{ i \phi} \rangle \langle \alpha e^{ i\phi} | \nonumber \\
& = e^{-|\alpha|^2} \sum_{n=0}^\infty \frac{|\alpha|^{2n}}{n!} | n \rangle \langle n | .
\label{diagonalalpha}
\end{align} 
where at steady-state the density matrix is completely diagonal.  As (\ref{intham}) only acts to produce a phase on off-diagonal components, on a diagonal state such as (\ref{diagonalalpha}) there is exactly zero effect.  In the case of Fig. \ref{fig4}(a), as the pump is always at a particular phase, hence there is an overall coherence that remains at steady-state.  But the more fragile phases due to the nonlinearity is completely removed, and no negativities remain.

\begin{figure}[t]
\includegraphics[width=\columnwidth]{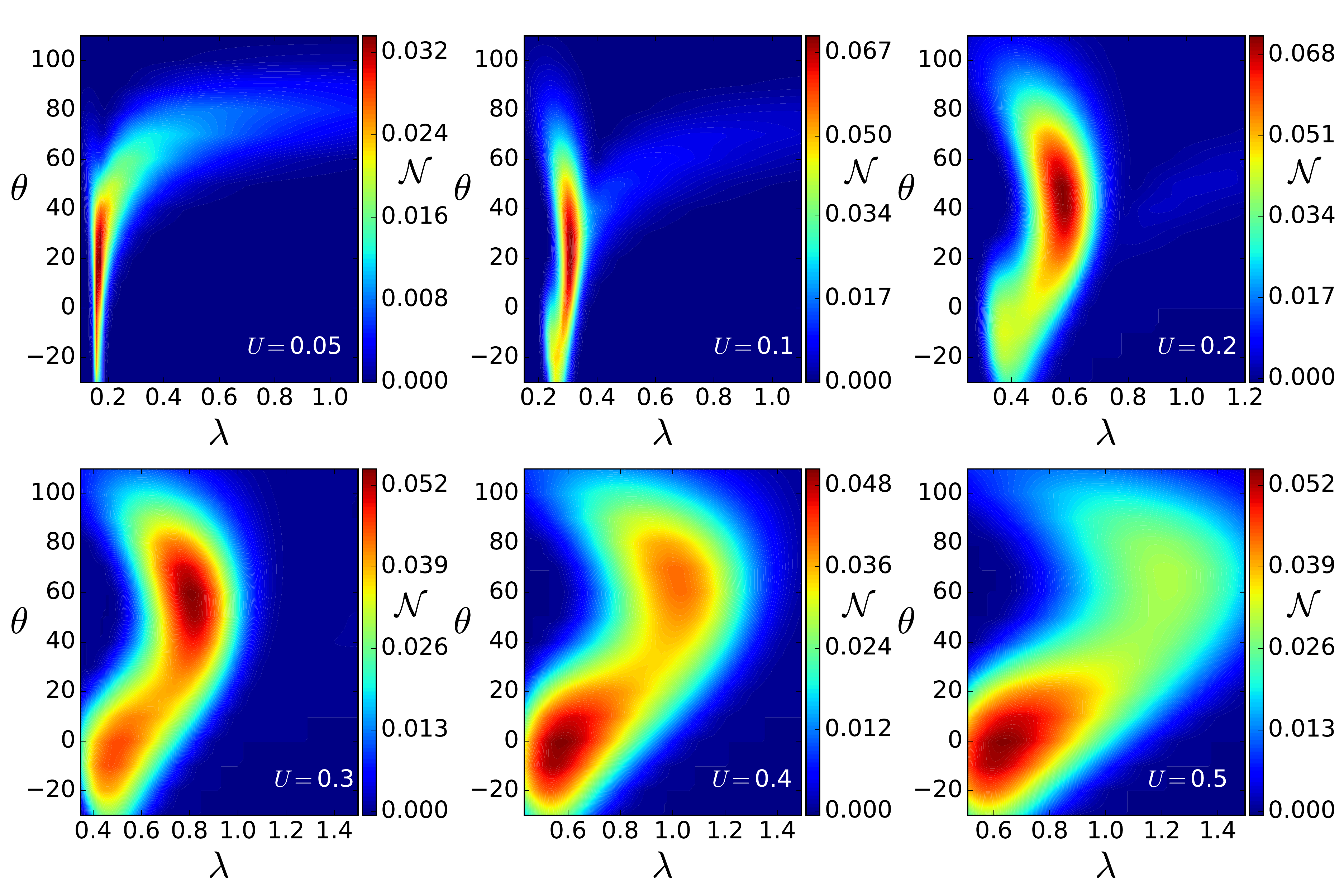}
\caption{\label{fig5} (Color online) Total integrated negativity $ {\cal N} $ in the space of the feedback parameter $ \lambda $ and pumping phase $ \theta $, for various nonlinear interaction coefficients $ U $. Common parameters are  $\Delta/\gamma=0$, $A/\gamma=3$, and $\eta =1.0$.} 
\end{figure}

Figure \ref{fig4}(b)(c) shows some typical results with feedback. We see that for much of the parameter range configurations with Wigner functions with a negative component can be created. For various parameters, the negative regions typically emerge as ripples of alternating positive and negative regions surrounding a primary amplitude peak at a fixed phase. Remarkably, steady-state negativities are achievable for relatively weak nonlinear coefficients in the region of $ U/\hbar \gamma = 0.05 $. The ratio of the nonlinear interaction to the cavity loss rate $ U/\hbar \gamma $  is one of the key parameters, as the nonlinearity is the source of the nonclassical light generation, and $ 1/ \gamma $ is the time that the cavity photons (or polaritons) have available before they decay.  Despite the decay rate being much faster than the interaction timescale, this is sufficient to produce significant Wigner negativities. We emphasize that the relevant timescales that the feedback should operate on should be the phase diffusion time, rather than the lifetime of the photons in the cavity.  For exciton-polaritons, this would correspond to the coherence time of the polaritons.  Although the lifetime of the polaritons is much shorter than the feedback response time currently, the coherence times of the polaritons are typically longer and thus can be closer to typical feedback timescales. 

In order to obtain convergence of our solutions a relatively large cutoff $ n_{\mbox{\tiny cut}} \approx 100 $ was required in our simulations for average photon numbers in the region of $ \langle n \rangle \sim 10$, where $ n = a^\dagger a $.  For this reason our simulations are generally for relatively small photon numbers $ \langle n \rangle \le 10$.  Within this regime, we found no difficulty in achieving steady-state negative Wigner functions for light with large amplitudes. We thus believe there is no inherent difficulty with creating bright amplitude light with negative Wigner components with the method.

\begin{figure}[t]
\includegraphics[width=\columnwidth]{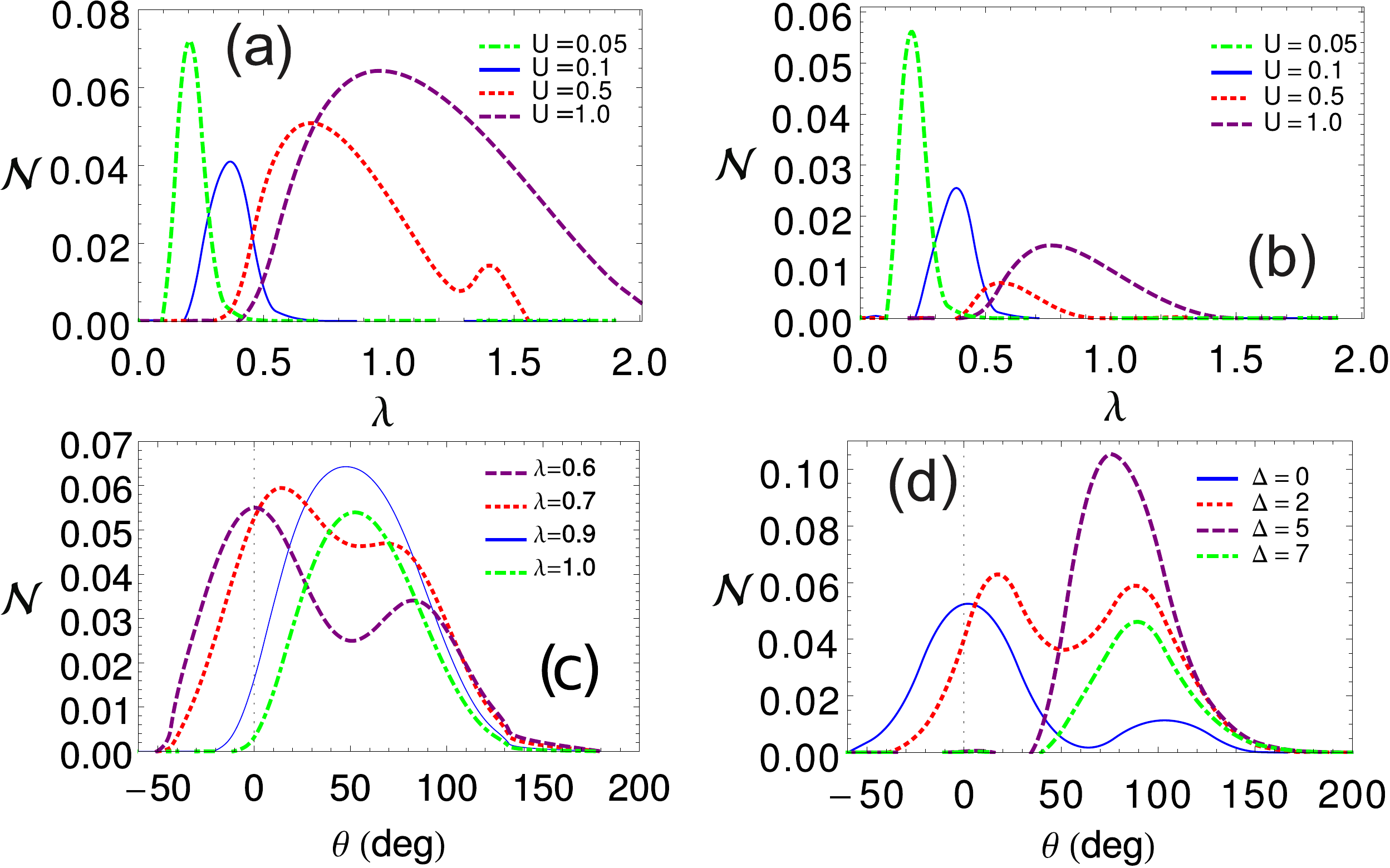}
\caption{\label{fig6}(Color online) Total integrated negativity $ {\cal N} $  as a function of various parameters. Parameters used are (a) $ \Delta/\gamma= 0 $, $ A/\gamma= 2 $, $\eta = 1 $, $\theta = 10^\circ $; (b) $ \Delta/\gamma= 0 $, $ A/\gamma= 2 $, $\eta = 0.7 $, $\theta = 10^\circ $;  (c) $ \Delta/\gamma= 0 $, $ A/\gamma= 2 $, $\eta = 1 $, $U/\hbar \gamma = 0.5 $; (d) $ \lambda/\gamma= 0.7 $, $ A/\gamma= 3 $, $\eta = 1.0 $, $U/\hbar \gamma = 0.5$.} 
\end{figure}

\section{Negativity of Wigner function}
\label{sec:negativity}

As a measure of the nonclassical nature of the light we calculate the total integrated negativity 
\begin{align}
{\cal N} \equiv \frac{1}{2} \int \Big( |W( \alpha ,\alpha^*)| - W( \alpha ,\alpha^*) \Big) d \alpha d \alpha^*   .      
\end{align}
In Fig. \ref{fig5} we show $ {\cal N} $ in the space of the feedback parameter $ \lambda/\gamma  $ and the pumping phase $ \theta $, for various nonlinear interactions $ U/\hbar \gamma $.  As suggested from Fig. \ref{fig4}(a) for all parameters with no feedback $\lambda=0 $ the negativity is zero. We see that there are optimal values of the feedback parameters for steady-state negativities to be achieved. We see that for larger interaction coefficients there is a broader range of parameters that allow for negativities. 
Looking at the region that produces the largest negativities, for larger interaction coefficients a stronger feedback coefficient is generally required.  This can be understood to originate from the the fact that the nonlinear coefficient itself acts to produce a ``twisting'' of the Wigner distribution which must be counteracted to produce a stable distribution.

We find that even for small $ U/\hbar \gamma =0.05 $, surprisingly it is possible to produce large amplitude light with negative Wigner function components. A comparison of the achievable negativities is shown in Fig. \ref{fig6}.  We see that the typical integrated negativity in the region of ${\cal N} \sim 0.05$ is achieved for all interaction coefficients. The price to be paid for this is the relatively narrow parameter region that this occurs in for small interaction parameters.  Interestingly, the effect of the nonlinear interactions is not to change the total negativity, but to make the parameter ranges for steady-state negativity broader.  We note that currently available cavities for exciton-polaritons are in the region of $ U/\hbar \gamma = 0.05 $, so in terms of the nonlinearity the technology is already in place. The more challenging aspects of the experiment are rather in the timescales of the phase diffusion (or coherence time) and the feedback circuitry that must be achieved.

\begin{figure}[t]
\includegraphics[width=\columnwidth]{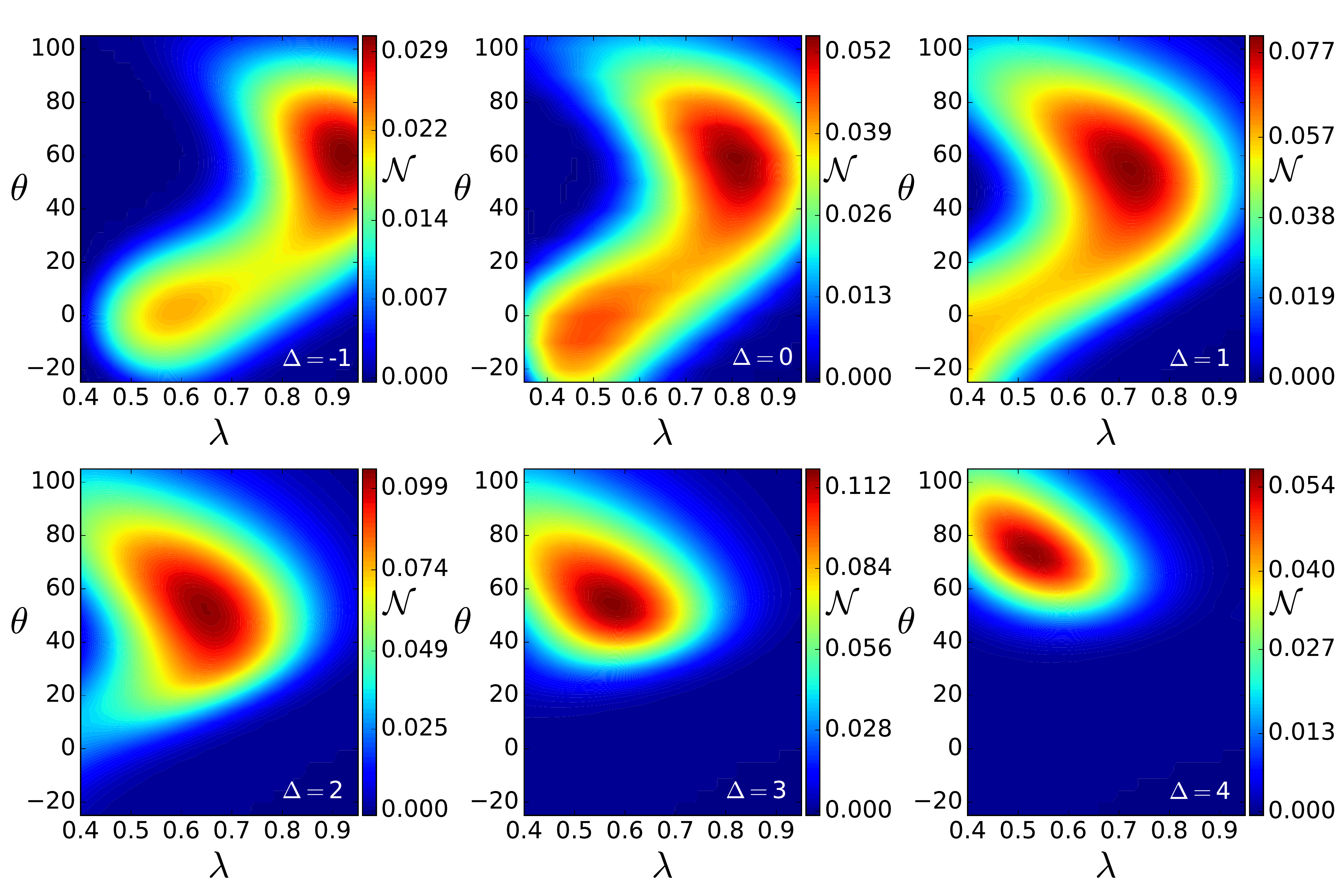}
\caption{\label{fig7}(Color online) Total integrated negativity $ {\cal N} $ in the space of the feedback parameter $ \lambda $ and pumping phase $ \theta $, for various detunings $ \Delta $. Common parameters are $U/\gamma\hbar=0.3$,  $A/\gamma=3$ and $\eta =1.0$. 
} 
\end{figure}

Turning to the variation of the total negativity with the pump laser phase angle $ \theta $, we see that the most effective pump is generally in the region of $ 0^\circ $  to $ 60^\circ $, although the precise optimum value depends on the feedback parameters (see Figs. \ref{fig5} and \ref{fig6}(c)). The effect of feedback control with a positive $ x $ homodyne current is to lock the output light at an angle of $\theta = 90^\circ$ \cite{wiseman93}.  Meanwhile 
the effect of the nonlinearity is to rotate the Wigner function in a counterclockwise direction.  We attribute the optimal angle to be in the observed range to be a combination of the two effects.  The negativity completely disappears for a range of pumping phases that are away from the locking region, thus it is crucial to pump with the correct phase. 

We also vary the detuning $ \Delta $ of the pump laser, which corresponds to adding a rotation to the phase of the condensate (see Figs. \ref{fig6}(d) and \ref{fig7}).  For exciton-polaritons, in practice only small detunings would be possible as the pump laser would need to be in resonance with the lower polariton branch.  We find that some higher peak negativities can be achieved with the detuning, usually at the cost of a narrower phase window in $ \theta $. In general however, the largest range of negativities occurs for 
$ \Delta =0 $. Therefore, some typical initial trial values for observing negativities would be $\theta \approx 30^\circ , \Delta = 0 ,  \lambda = 0.3 $.  These are not the optimal parameters but they may be used as a starting point for further refinement.

Finally, Fig. \ref{fig8} shows the effect of imperfect effective detection efficiency.  As discussed in the Appendix, the effective detector efficiency is simply the product of the bare photon detector efficiency, and the reflectivity of the beam splitter as given in Fig. \ref{fig1}, which feeds back part of the emerging light to the homodyne measurement.  We see that the effect of $ \eta $ is generally to reduce the total negativity, as expected.  We observe from Fig. \ref{fig6}(b) that interestingly, the effect of finite detection efficiency is less for smaller nonlinearities, which is advantageous from a experimental point of view. It is also interesting to observe that there is a threshold effective detection efficiency to observe a non-zero negativity. For typical values this is of the order of $ \eta \sim 0.5 $ hence is not unreasonable from an experimental point of view.

\begin{figure}[t]
\includegraphics[width=\columnwidth]{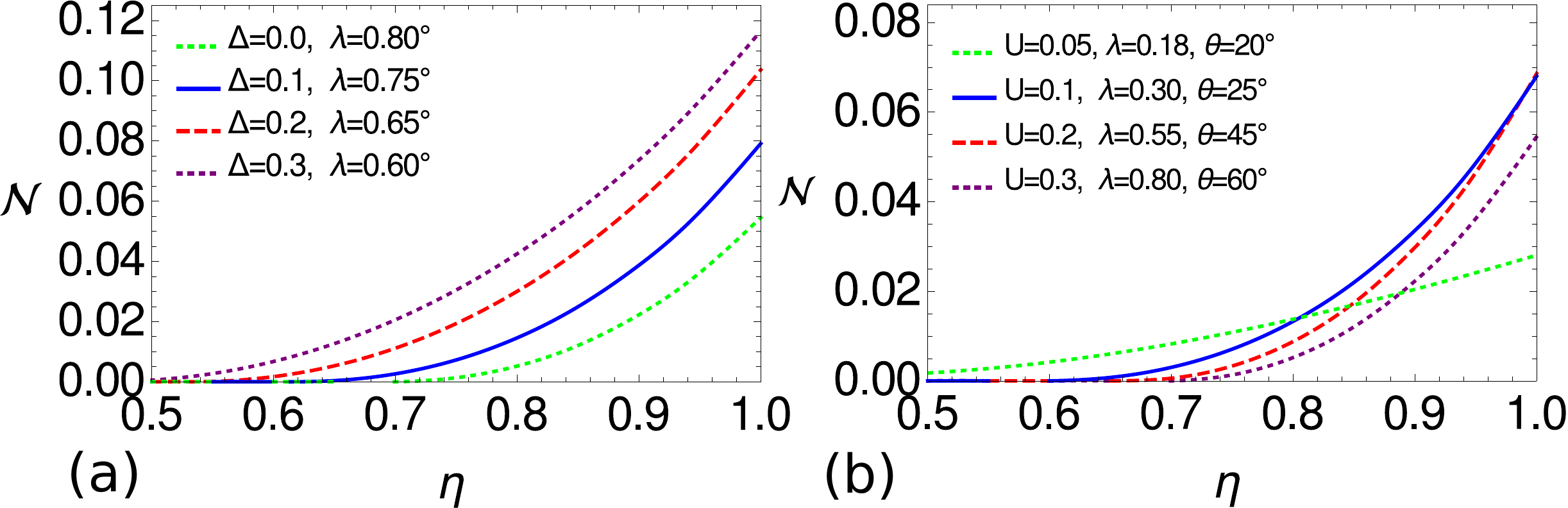}
\caption{\label{fig8}(Color online)   Total integrated negativity $ {\cal N} $ as a function of the effective detection efficiency $\eta$ for different parameters as marked. The common used parameters are (a)  $A/\gamma=3$, $U/\gamma\hbar=0.3$, $\theta =55^\circ$  and (b) $A/\gamma=3$, $\Delta=0.0$.
}
\end{figure}

\section{Summary and conclusions}
\label{sec:conc}

We have proposed a scheme for generating steady-state nonclassical light with a negative Wigner function component using a combination of a nonlinear medium in a cavity and feedback. This generalizes our scheme presented in  Ref. \cite{byrnes13}, which was for the more specific case of polariton relaxation in a microcavity.  In principle, this could be used in any configuration where a nonlinearity is present within a cavity, if it can be controlled via feedback.  The results in this paper are analogous to those presented in Ref. \cite{wiseman93}, where the aim is to prepare squeezed states via quantum feedback.  In this paper, with the help of a nonlinear medium, light with negative Wigner functions can be stabilized at steady-state.  Without the feedback, even arbitrarily strong nonlinear media are unable to produce Wigner negativity at steady-state, due to an effective phase diffusion of the cavity.  We showed for particular experimental parameters derived from exciton-polaritons that light with negative Wigner functions can be generated at steady-state with current experimental parameters.  

In the context of exciton-polaritons, our scheme improves upon our previous scheme of Ref. \cite{byrnes13} in several ways.  First, the pumping mechanism for producing the polaritons is simplified by considering direct coherent excitation of the polariton condensate. Another improvement is the degree of the nonclassicality of the emerging light. Previously, the total negativity was of the level of $ {\cal N} \sim 0.01 $ for typical experimental parameters; our current results improve upon this by an order of magnitude for SWG cavities that are available today. The stabilization of the phase of the polaritons is required for both approaches, in our previous study an injection locking technique was used, in contrast to the feedback approach employed here. From our present results the feedback appears to be a more powerful technique for achieving Wigner negativities.  While it is technically more demanding than the relatively simple injection locking method, such feedback has already been achieved in the context of lasers \cite{huang08}, hence the extension to exciton-polaritons should be relatively straightforward. We anticipate the most challenging aspect of the experiment will be to match the feedback response timescale to the coherence time of the polaritons. 

If experimentally realized such a device would produce light with a negative Wigner function continuously. We note that our simulations simulate the Wigner functions for the photons inside the cavity.  Calculating the light that emerges in the free mode beyond the beamsplitter (mode $a' $ in Fig. \ref{fig9}) requires a more involved calculation which we leave as future work.  Producing light with Wigner negativity in the free mode  would constitute an unusual type of non-classical laser that produces coherent non-Gaussian light. Current methods of producing such light, such as photon subtraction, are inherently probabilistic and would be distinct in the sense that this is a continuous light source. This may be useful for applications in continuous variables quantum information, where it is known that non-Gaussian light is required for universal quantum computation \cite{lloyd99,bartlett02,sefi11,broecker95,mandilara09,mari12}. 

\acknowledgments

T. B. thanks Sven H{\"o}fling and Howard Wiseman for discussions. This work is supported by JSPS KAKENHI Grant Number 26790061, the Shanghai Research Challenge Fund, New York University Global Seed Grants for Collaborative Research, National Natural Science Foundation of China (NSFC) grant 61571301, the Thousand Talents Program for Distinguished Young Scholars, and the NSFC Research Fund for International Young Scientists. H. D. acknowledges support by the National Science Foundation (DMR 1150593) and the Air Force Office of Scientific Research (FA9550-15-1-0240).

\appendix 

\section{Effective detector efficiency}
\label{sec:appen1}

\begin{figure}[b]
\includegraphics[width=\columnwidth]{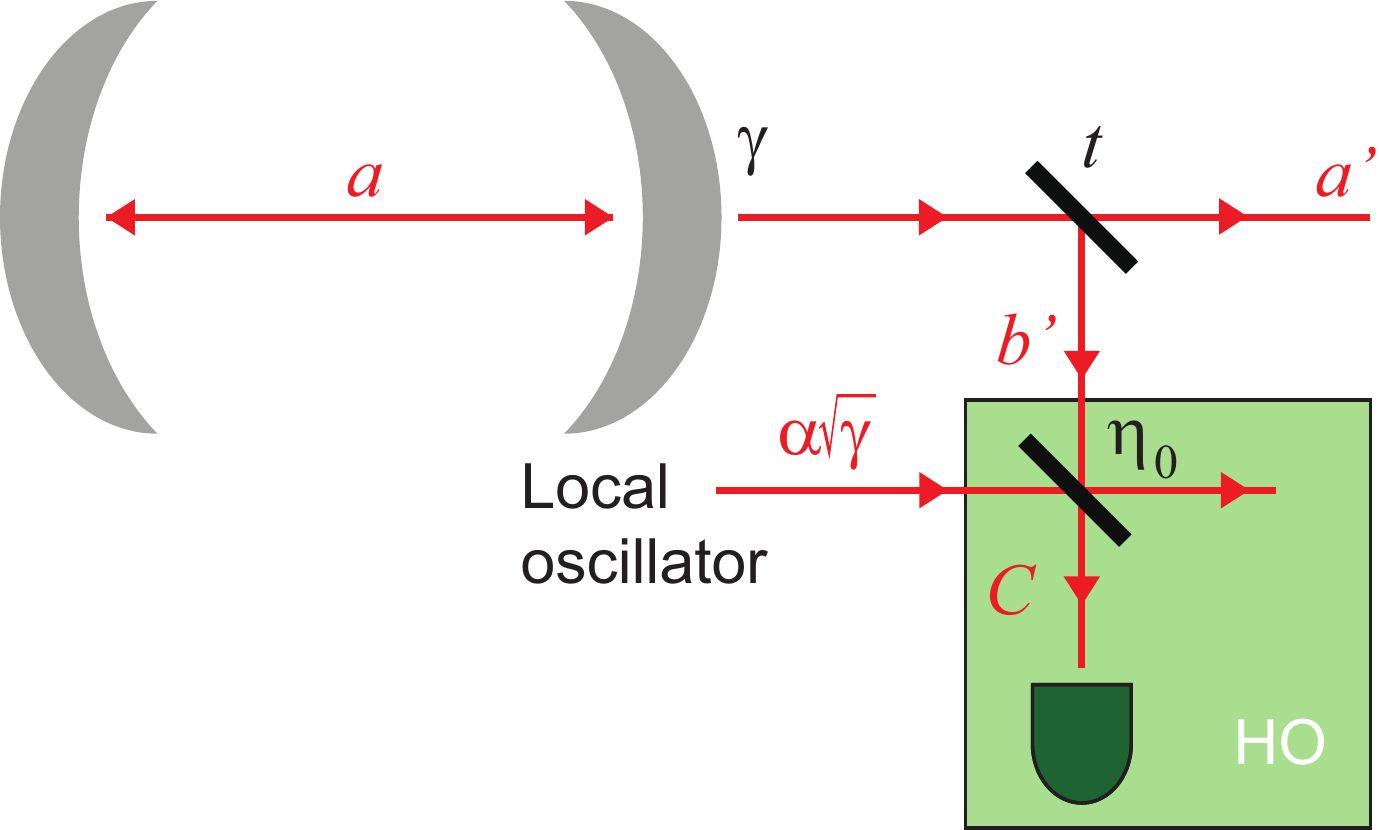}
\caption{\label{fig9}(Color online) Modes involved in proposed experimental system.  The first beam splitter with transmittivity $ t $ acts to divert a portion $ r = 1- t $ of the light to be used for the feedback system.  The homodyne system has a finite detector efficiency $ \eta_0 $ which can be equivalently considered to be the beam splitter transmittivity \cite{wiseman93}.  }
\end{figure}

Here we show that the effective detector efficiency $ \eta $ has contributions from both the bare homodyne detector efficiency $ \eta_0 $ and the beamsplitter to divide the light between the output and the homodyne detection as shown in Fig. \ref{fig1}. Figure \ref{fig9} shows a schematic diagram showing the various modes that are present in the experimental system.  This is an adaptation of the same argument as shown in Ref. \cite{wiseman93}, where the feedback equations for homodyne detection are derived.  First, the mode emerging from the cavity is damped by a factor equal to the cavity loss, and enters a beam splitter with transmittivity $ t $ (or reflectivity $ r = 1- t $).  The emerging modes are thus \cite{PhysRevA.47.642}
\begin{align}
b' & = \sqrt{\gamma} \left( \sqrt{r} a   + i \sqrt{1-r} b \right)
\label{firstbs} 
\end{align}
where $b$ is the input of the other port of the beam splitter, which in this case is the vacuum. The mode $ b'$ now acts as the input to another beam splitter, this time combined with a local oscillator of amplitude $\alpha \sqrt{\gamma} $ inside the homodyne detector, with efficiency $\eta_0$.  The mode $C$ is given by 
\begin{align}
C = \sqrt{\eta_0}  b'  + i \alpha  \sqrt{\gamma} \sqrt{1-\eta_0} .  
\end{align}
Combining this with (\ref{firstbs}), we obtain
\begin{align}
 C &= \sqrt{\gamma r \eta_0}a  + i \sqrt{\gamma(1-r)\eta_0  }  b    +  i\sqrt{(1-\eta_0)\gamma} \alpha    \\
  & \approx \sqrt{\gamma} \left( \sqrt{\eta_0 r}  a  + i \sqrt{1-\eta_0} \alpha \right)
\end{align}
where in the second line we assumed that mode $ b $ is the vacuum, and in combination with the local oscillator the vacuum noise can be neglected. We thus see by comparison to Eq. (4.2) in Ref. \cite{wiseman93} that the effect of the first beam splitter is to modify the effective detection efficiency of the homodyne detection, giving 
\begin{align}
\eta = \eta_0 r , 
\end{align}
where $ \eta $ is the effective detection efficiency as discussed in the main text. In this sense, the beam splitter that divides the light between the feedback and output plays an active role in determining the system dynamics, by contributing to a reduced effective detector efficiency.

\bibliographystyle{apsrev}
\bibliography{NIIReferences}

\end{document}